\documentclass[english,aps, superscriptaddress, floatfix, 12pt]{revtex4}
\usepackage[T1]{fontenc}
\usepackage[latin1]{inputenc}
\usepackage{graphicx}
\usepackage{amssymb}
\usepackage{epsfig}
\usepackage{amsmath}
\usepackage{psfrag}
\usepackage[active]{srcltx}
\usepackage{babel}
\makeatother   
\begin{document}

\title{Quantifying zigzag motion of quarks}
\author{D. Antonov, J.E.F.T. Ribeiro\\
{\it Departamento de F\'isica and Centro de F\'isica das Interac\c{c}\~oes Fundamentais,}\\ 
{\it Instituto Superior T\'ecnico, UT Lisboa,
Av. Rovisco Pais, 1049-001 Lisboa, Portugal}}

\begin{abstract}
The quark condensate is calculated in terms of the effective string tension and the constituent quark mass.
For 3 colors and 2 light flavors, the constituent mass is bounded from below by
the value of 460~MeV. This value is only accessible when the string tension decreases linearly with the 
Schwinger proper time. For this reason, the Hausdorff dimension of a
light-quark trajectory is equal to 4, indicating that these trajectories are similar 
to branched polymers, which 
can describe a weak first-order deconfinement phase transition in SU(3) Yang--Mills theory. 
Using this indication, we develop a gluon-chain model based on such trajectories.
\end{abstract}

\maketitle

\section{Introduction}

\noindent
At zero temperature, the classical action of QCD with $N_{\rm f}$ massless quark flavors possesses the
so-called chiral SU$_{\rm L}(N_{\rm f})\times$SU$_{\rm R}(N_{\rm f})$ symmetry. On the quantum
level, this symmetry is spontaneously broken, and the order parameter for the symmetry breaking is
the chiral condensate $\bigl<\bar\psi\psi\bigr>$. Together with confinement, characterized by the
gluon condensate $\bigl<g^2(F_{\mu\nu}^a)^2\bigr>$, chiral-symmetry breaking is one of the two
most fundamental nonperturbative phenomena in QCD. We notice that the problem of interrelation
between the chiral-symmetry breaking and confinement was already considered 
in Ref.~\cite{ne}. In those papers, within a Nambu--Jona-Lasinio--type model with confinement
incorporated, a proportionality of the two condensates was found. Here, we revisit this problem 
in a different framework. To this end, we use the definition of the
quark condensate $\bigl<\bar\psi\psi\bigr>$ in terms of the one-loop effective action,
\begin{equation}
\label{one}
\bigl<\bar\psi\psi\bigr>=-\frac{\partial}{\partial m}\bigl<\Gamma[A_\mu^a]\bigr>.
\end{equation}
By virtue of the world-line representation for $\Gamma[A_\mu^a]$ (cf. Ref.~\cite{WL}), such an
approach allows us to relate the condensation of a quark to the zigzag behavior of its trajectories.
Quark condensation requires a 
$1/\sqrt{s}$ asymptotic behavior of the integral over the trajectories at 
large Schwinger proper times~$s$~\cite{bc}. To what extent this zigzagness is needed to provide 
quark condensation and how to reconcile it with quark confinement are the questions addressed 
in this paper.

Being motivated by the above-mentioned proportionality~\cite{ne} between the chiral 
and the gluon condensates, we calculate $\bigl<\Gamma[A_\mu^a]\bigr>$ by imposing a confining area-law 
behavior of the quark's Wilson loop. To this end, we introduce a parametrization of the minimal area
rendering the path integral, which represents $\bigl<\Gamma[A_\mu^a]\bigr>$, analytically calculable.
Then, the $1/\sqrt{s}$ asymptotic behavior of this path integral in the chiral limit, together 
with the known expression for $\bigl<\bar\psi\psi\bigr>$ in the heavy-quark limit, 
leads to the $1/s$-scaling of the
effective string tension. We will show that this scaling corresponds to quark 
trajectories with the Hausdorff dimension equal to 4. Such trajectories, which are crumpled more
severely than the ordinary Brownian random walk, are characteristic of the so-called branched polymers. To illustrate the
physical significance of this class of trajectories, we have developed a model of the deconfinement phase transition,
 which is based on a gluon chain extending
along such a trajectory. It turns out that this model describes a weak first-order phase
transition, unlike its counterpart based on Brownian random walks, which describes a second-order 
phase transition~\cite{adp}. As such, it is an interesting example of a 
QCD-inspired model, which can analytically 
describe the deconfinement phase transition in SU(3) Yang--Mills theory.

The outline of the paper is as follows. In the next section, we develop a parametrization of the
minimal area in terms of the contour of a Wilson loop and apply it to the calculation of the
quark condensate. Starting with the purely exponential dependence of a Wilson loop on the minimal area, we then 
proceed, motivated by the heavy-quark limit, to more complicated ans\"atze. These ans\"atze, 
involving a pre-exponential area-dependence, allow us to recover within the same formalism the 
``area-squared'' law of small Wilson loops.
In this way, we find, within the present approach, the lowest possible bound on the value of the constituent 
quark mass, and demonstrate, both analytically and numerically, the $1/s$-scaling of the effective 
string tension. In section~III, we argue that this scaling corresponds to quark trajectories with 
Hausdorff dimension 4, and use 
such trajectories to develop a model of the deconfinement phase transition in SU(3)
Yang--Mills theory. In section~IV, we summarize the main results of the paper.
In Appendix~A, we check that the known result for the heavy-quark condensate is correctly reproduced 
by small Wilson loops, which depend quadratically on the minimal area. We furthermore show there 
that the ansatz for the Wilson loop involving the pre-exponential area-dependence can be 
used to fit the quadratic area-dependence of small Wilson loops. In Appendix~B, we illustrate for the simpler case of heavy quarks, how to get, within our approach, an equivalent confining Nambu--Jona-Lasinio--type model.
\section{Quark condensate from the minimal-area law}

\noindent
We use Eq.~(\ref{one}), where the averaged  
one-loop effective action reads~\cite{WL} 
$$\left<\Gamma[A_\mu^a]\right>=-(2S+1)N_{\rm f}\int_0^\infty
\frac{ds}{s}{\rm e}^{-m^2s}
\int_{P}^{}{\cal D}z_\mu \int_A^{}{\cal D}\psi_\mu
{\rm e}^{-\int_0^s d\tau\left(\frac14\dot z_\mu^2+\frac12\psi_\mu\dot\psi_\mu\right)}\times$$
\begin{equation}
\label{ttuu}
\times\left\{\left<{\rm tr}{\,}{\cal P}{\,}
\exp\left[ig\int_0^s d\tau T^a \left(A_\mu^a\dot z_\mu-\psi_\mu\psi_\nu F_{\mu\nu}^a\right)
\right]\right>-N_c\right\}.
\end{equation}
In this formula, $P$ and $A$ stand, respectively, for the periodic and the antiperiodic boundary conditions, 
so that
$\int_P^{}\equiv\int_{z_\mu(s)=z_\mu(0)}^{}$, $\int_A^{}\equiv\int_{\psi_\mu(s)=-\psi_\mu(0)}^{}$. The  
trajectories $z_\mu(\tau)$ obey the equation
\begin{equation}
\label{peri}
\int_0^s d\tau z_\mu(\tau)=0.
\end{equation}
Equation~(\ref{peri}) means that the 
center of a trajectory is the origin, i.e. the factor of volume associated with the translation of 
a trajectory as a whole is already divided out.
In Eq.~(\ref{ttuu}), $T^a$ is a generator of the SU($N_c$)-group in the fundamental representation, $S$ is the 
spin of the fermion, which propagates along the loop (i.e., for a quark, $S=1/2$), and $N_{\rm f}$ 
is the number of light-quark flavors. Furthermore, 
since the quark condensation can occur only due to the gauge fields, we have subtracted the free part of the 
effective action, so that $\bigl<\Gamma[0]\bigr>=0$.

The field-strength tensor entering the spin term $\sim\psi_\mu\psi_\nu F_{\mu\nu}^a$ 
can be recovered by means of 
the area-derivative operator $\frac{\delta}{\delta s_{\mu\nu}}$
acting on the Wilson loop~\cite{mm}. This fact allows us to reduce
the gauge-field dependence of Eq.~(\ref{ttuu}) to that of the Wilson loop,
$$\bigl<W[z_\mu]\bigr>=\left<{\rm tr}{\,}{\cal P}{\,}\exp\left(ig\int_0^s d\tau T^aA_\mu^a\dot z_\mu\right)\right>,$$
as follows:
$$\bigl<\Gamma[A_\mu^a]\bigr>=-2N_{\rm f}\int_0^\infty\frac{ds}{s}{\rm e}^{-m^2s}
\int_{P}^{}{\cal D}z_\mu \int_{A}^{}
{\cal D}\psi_\mu {\rm e}^{-\int_0^s d\tau\left(\frac14\dot z_\mu^2+\frac12\psi_\mu\dot\psi_\mu\right)}\times$$
\begin{equation}
\label{effA}
\times\left\{\exp\left[-2\int_0^s d\tau\psi_\mu\psi_\nu\frac{\delta}{\delta s_{\mu\nu}(z(\tau))}\right]
\bigl<W[z_\mu]\bigr>-N_c\right\}.
\end{equation}
We proceed now with the construction of the minimal area of the Wilson loop $\bigl<W[z_\mu]\bigr>$ in a such a way as to render the path integral analytically calculable.
\subsection{A simple ansatz for the minimal area of the Wilson loop}

\noindent
We ignore perturbative interactions of the quarks, which in any case do not affect the {\it nonperturbative} quark condensate. Hence, let us start by introducing a simple ansatz for the Wilson loop, such as the
one given by minimal-area law:
\begin{equation}
\label{ans}
\bigl<W[z_\mu]\bigr>=N_c\cdot {\rm e}^{-\sigma(s)\cdot S_{\rm min}[z_\mu]}.
\end{equation}
Here, $\sigma(s)$ is some proper-time dependent string tension of a quark, and $S_{\rm min}[z_\mu]$ is the 
area of the minimal surface bounded by the quark trajectory $z_\mu(\tau)$.
In the purely spatial 3-dimensional case, the minimal surface can naturally be 
built out of infinitely thin segments, so that its area reads 
$S_{3{\rm d}}=\frac12\int_0^s d\tau|{\bf z}\times\dot{\bf z}|$~\cite{aps}. Clearly, this is just an area of the 
surface formed by a rotating {\it rod} of a variable length. As such, this area is minimal from among 
the areas of all possible surfaces which can be formed by rotating curved lines.
Furthermore, a 4-dimensional generalization of $S_{3{\rm d}}$ is given by 
$$S_{4{\rm d}}=\frac{1}{2\sqrt{2}}\int_0^s d\tau|\varepsilon_{\mu\nu\lambda\rho}z_\lambda\dot z_\rho|.$$
The constant in front of the integral in this formula is fixed in such a way as to have 
$S_{4{\rm d}}$ going over to 
$S_{3{\rm d}}$ when $z_4=\dot z_4\equiv 0$. Notice also that, owing to condition~(\ref{peri}), the 
surface pieces forming the macroscopic surface merge at the origin. For this reason, from among 
all the areas of cone-shaped surfaces with the 
boundary $z_\mu(\tau)$, the minimal one is automatically given by $S_{4{\rm d}}$.

The functional $S_{4{\rm d}}[z_\mu]$ can furthermore be minorated as follows. Denoting 
$$\varepsilon_{\mu\nu\lambda\rho}z_\lambda\dot z_\rho\equiv\sigma_{\mu\nu},$$ 
we have 
$$\left|\int_0^s d\tau\sigma_{\mu\nu}\right|=\sqrt{2\cdot\sum\limits_{\mu<\nu}^{}\Biggl(
\int_0^s d\tau\sigma_{\mu\nu}\Biggr)^2}\le\sqrt{2}\cdot\sum\limits_{\mu<\nu}^{}\left|
\int_0^s d\tau\sigma_{\mu\nu}\right|\le\sqrt{2}\int_0^s d\tau\Biggl(\sum\limits_{\mu<\nu}^{}|\sigma_{\mu\nu}|
\Biggr).$$
Next, applying the Cauchy--Schwarz inequality in the form $\sum\limits_{i=1}^n|a_i|\le\sqrt{n\cdot
\sum\limits_{i=1}^n a_i^2}$, where in our case $n=6$, we get 
$$\sqrt{2}\int_0^s d\tau\Biggl(\sum\limits_{\mu<\nu}^{}|\sigma_{\mu\nu}|
\Biggr)\le \sqrt{12}\int_0^s d\tau\sqrt{\sum\limits_{\mu<\nu}^{}\sigma_{\mu\nu}^2}=
\sqrt{6}\int_0^s d\tau|\sigma_{\mu\nu}|.$$
Therefore,
\begin{equation}
\label{re18}
S_{4{\rm d}}=\frac{1}{2\sqrt{2}}\int_0^s d\tau|\sigma_{\mu\nu}|\ge\frac{1}{2\sqrt{12}}
\left|\int_0^s d\tau\sigma_{\mu\nu}\right|.
\end{equation}
In what follows, we assume that $S_{4{\rm d}}^{\rm min}$ is provided by 
the right-hand side of this inequality. Denoting the integrated surface element and the rescaled 
string tension respectively as 
$$\Sigma_{\mu\nu}\equiv\int_0^s d\tau\sigma_{\mu\nu}~~~~ {\rm and}~~~~ \tilde\sigma\equiv\tilde\sigma(s)=
\frac{\sigma(s)}{2\sqrt{12}},$$
we can write the Wilson loop, Eq.~(\ref{ans}), in the simple form
\begin{equation}
\label{ans12}
\bigl<W[z_\mu]\bigr>=N_c\cdot {\rm e}^{-\tilde\sigma|\Sigma_{\mu\nu}|}.
\end{equation} 
We will show that the $1/\sqrt{s}$ asymptotic behavior at large $s$ of the path integral in
Eq.~(\ref{ttuu}) in the chiral limit leads to a calculable $s$-dependence 
of the {\it effective string tension} $\tilde\sigma$.

By means of an auxiliary integration, we can further proceed from $|\Sigma_{\mu\nu}|$ to $\Sigma_{\mu\nu}^2$
in the exponent in Eq.~(\ref{ans12}):
$$\bigl<W[z_\mu]\bigr>=N_c\int_0^\infty\frac{d\lambda}{\sqrt{\pi\lambda}}{\rm e}^{-\lambda-
\frac{\tilde\sigma^2}{4\lambda}\Sigma_{\mu\nu}^2}.$$
We now apply the Hubbard--Stratonovich trick, which amounts to introducing an auxiliary constant Abelian 
field $B_{\mu\nu}$:
\begin{equation}
\label{HS18}
{\rm e}^{-\frac{\tilde\sigma^2}{4\lambda}\Sigma_{\mu\nu}^2}=\left(\frac{\lambda}{2\pi\tilde\sigma^2}
\right)^3\cdot\left(\prod\limits_{\mu<\nu}^{}\int_{-\infty}^{+\infty}dB_{\mu\nu}\right)
{\rm e}^{-\frac{\lambda B_{\mu\nu}^2}{4\tilde\sigma^2}-\frac{i}{2}B_{\mu\nu}\Sigma_{\mu\nu}}.
\end{equation}
The trajectory-dependence appears now in the form ${\rm e}^{-\frac{i}{2}B_{\mu\nu}\Sigma_{\mu\nu}}$, 
which can be recognized as a Wilson loop of an electron of charge $(-1)$ in the electromagnetic field with the 
strength tensor
\begin{equation}
\label{jj7}
{\cal F}_{\mu\nu}\equiv\varepsilon_{\mu\nu\lambda\rho}B_{\lambda\rho}.
\end{equation}
The $\lambda$-integration can be performed analytically, 
$$\int_0^\infty d\lambda\lambda^{5/2}{\rm e}^{-\lambda\left(1+\frac{B_{\mu\nu}^2}{4\tilde\sigma^2}
\right)}=\frac{15\sqrt{\pi}}{8}\cdot\frac{1}{\left(1+\frac{B_{\mu\nu}^2}{4\tilde\sigma^2}
\right)^{7/2}},$$
to arrive at the following Abelian-like form for the Wilson loop:
\begin{equation}
\label{gg77}
\bigl<W[z_\mu]\bigr>=N_c\cdot\frac{15}{8}\cdot\frac{1}{(2\pi\tilde\sigma^2)^3}\cdot
\left(\prod\limits_{\mu<\nu}^{}\int_{-\infty}^{+\infty}dB_{\mu\nu}\right)
\frac{{\rm e}^{-\frac{i}{2}B_{\mu\nu}\Sigma_{\mu\nu}}}{\left(1+\frac{B_{\mu\nu}^2}{4\tilde\sigma^2}
\right)^{7/2}}.
\end{equation}

\subsection{Calculation of the quark condensate}

\noindent
We are left to calculate the effective action~(\ref{effA}) and the quark condensate~(\ref{one}), by using the 
obtained parametrization of the Wilson loop, Eq.~(\ref{gg77}). To this end, we notice that
the term $B_{\mu\nu}\Sigma_{\mu\nu}={\cal F}_{\mu\nu}\int_0^s d\tau z_\mu\dot z_\nu$, in the exponent 
of Eq.~(\ref{gg77}), can be written with the help of the (Abelian) Stokes' theorem as 
$${\cal F}_{\mu\nu}\int_0^s d\tau z_\mu\dot z_\nu = \int d s_{\mu\nu}{\cal F}_{\mu\nu},$$
where the integration on the right-hand side goes over an arbitrary surface bounded by the contour $z_\mu(\tau)$.
Therefore, when plugging Eq.~(\ref{gg77}) into Eq.~(\ref{effA}), we have 
$$\exp\left[-2\int_0^s d\tau\psi_\mu\psi_\nu\frac{\delta}{\delta s_{\mu\nu}(z(\tau))}\right]
{\rm e}^{-\frac{i}{2}B_{\mu\nu}\Sigma_{\mu\nu}}=\exp\left[-\int_0^s d\tau\left(\frac{i}{2}{\cal F}_{\mu\nu}
z_\mu\dot z_\nu-i{\cal F}_{\mu\nu}\psi_\mu \psi_\nu\right)\right].$$
Accordingly, the path integral in Eq.~(\ref{effA}) becomes that of an electron of charge $(-1)$ in a constant 
electromagnetic field:
$$\int_{P}^{}{\cal D}z_\mu \int_{A}^{}
{\cal D}\psi_\mu {\rm e}^{-\int_0^s d\tau\left(\frac14\dot z_\mu^2+\frac12\psi_\mu\dot\psi_\mu\right)}
\left\{\exp\left[-2\int_0^s d\tau\psi_\mu\psi_\nu\frac{\delta}{\delta s_{\mu\nu}(z(\tau))}\right]
{\rm e}^{-\frac{i}{2}B_{\mu\nu}\Sigma_{\mu\nu}}-1\right\}=$$
\begin{equation}
\label{eh}
=\int_{P}^{}{\cal D}z_\mu \int_{A}^{}{\cal D}\psi_\mu\exp\left[-\int_0^s d\tau\left(
\frac14\dot z_\mu^2+\frac12\psi_\mu\dot\psi_\mu+\frac{i}{2}{\cal F}_{\mu\nu}z_\mu\dot z_\nu-
i{\cal F}_{\mu\nu}\psi_\mu\psi_\nu\right)\right]-\frac{1}{(4\pi s)^2}.
\end{equation}
The result of integration is given by the so-called Euler--Heisenberg--Schwinger Lagrangian~\cite{EH}
(for reviews see~\cite{WL, fr}), so that 
\begin{equation}
\label{kk99}
{\rm Eq.}~(11){\,}={\,}\frac{1}{(4\pi s)^2}\left[s^2ab\cot(as)\coth(bs)-1\right].
\end{equation}
In this formula,
$$a^2=\frac12\left[{\bf E}^2-{\bf H}^2+\sqrt{({\bf E}^2-{\bf H}^2)^2+4({\bf E}\cdot{\bf H})^2}\right]$$ 
and
$$b^2=\frac12\left[{\bf H}^2-{\bf E}^2+\sqrt{({\bf E}^2-{\bf H}^2)^2+4({\bf E}\cdot{\bf H})^2}\right]$$
are related to the two invariants of the auxiliary electromagnetic field ${\cal F}_{\mu\nu}$ as 
$a^2-b^2={\bf E}^2-{\bf H}^2$, $a^2b^2=({\bf E}\cdot{\bf H})^2$. Equation~(\ref{kk99}) is a 
compact-form representation of the fermionic determinant, which can be expanded in the number 
of external lines of the electromagnetic field. Retaining in such an expansion the leading term, which 
corresponds to the one-loop diagram with only two external lines, we have 
$${\rm Eq.}~(11){\,}={\,}\frac{1}{(4\pi s)^2}\left[\frac{s^2}{3}(b^2-a^2)+{\cal O}
\bigl(s^4|{\cal F}_{\mu\nu}|^4\bigr)\right]=\frac{1}{(4\pi s)^2}\left[\frac{s^2}{3}
\sum\limits_{\mu<\nu}^{}{\cal F}_{\mu\nu}^2+{\cal O}
\bigl(s^4|{\cal F}_{\mu\nu}|^4\bigr)\right].$$
Returning to the $B_{\mu\nu}$-field, which is related to the ${\cal F}_{\mu\nu}$-field through Eq.~(\ref{jj7}),
and neglecting the terms ${\cal O}\bigl(s^4|B_{\mu\nu}|^4\bigr)$, we can write
$${\rm Eq.}~(11){\,}\simeq\frac{1}{12\pi^2}\sum\limits_{\mu<\nu}^{}B_{\mu\nu}^2.$$
This expression, along with Eq.~(\ref{gg77}), yields for the one-loop effective action, Eq.~(\ref{effA}):
$$\bigl<\Gamma[A_\mu^a]\bigr>=-2N_{\rm f}N_c\cdot\frac{1}{12\pi^2}\cdot\frac{15}{8}\cdot
\frac{1}{(2\pi)^3}\int_0^\infty\frac{ds}{s}\cdot\frac{{\rm e}^{-m^2s}}{\tilde\sigma^6}
\left(\prod\limits_{\mu<\nu}^{}
\int\limits_{-1/s}^{1/s} dB_{\mu\nu}\right)\frac{\vec B{\,}^2}{\left(1+\frac{\vec B{\,}^2}{2\tilde\sigma^2}
\right)^{7/2}}.$$
In this formula, $\vec B$ is a 6-vector,
$$\vec B=(B_{12},B_{13},B_{14},B_{23},B_{24},B_{34}),~~~~ {\rm so}~~ {\rm that}~~ 
\sum\limits_{\mu<\nu}B_{\mu\nu}^2=\vec B{\,}^2.$$
Furthermore, the neglection of the terms ${\cal O}\bigl(s^4|B_{\mu\nu}|^4\bigr)$ leads to the corresponding 
restriction of the integration range. 

\begin{figure}
\epsfig{file=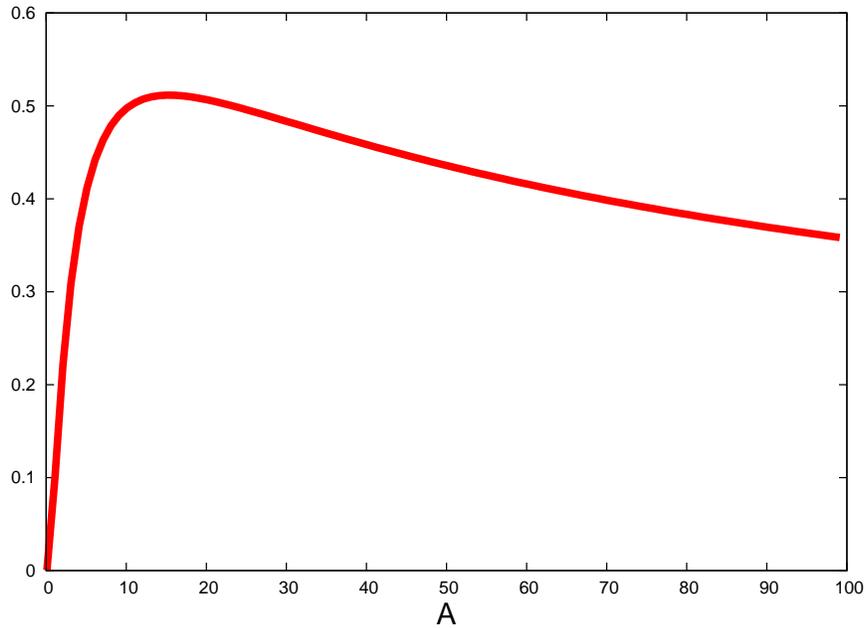, width=120mm}
\caption{\label{1}The function $G[A]=\frac{f[A]}{A}$ for $A\in(0,100)$.}
\end{figure}

\begin{figure}
\epsfig{file=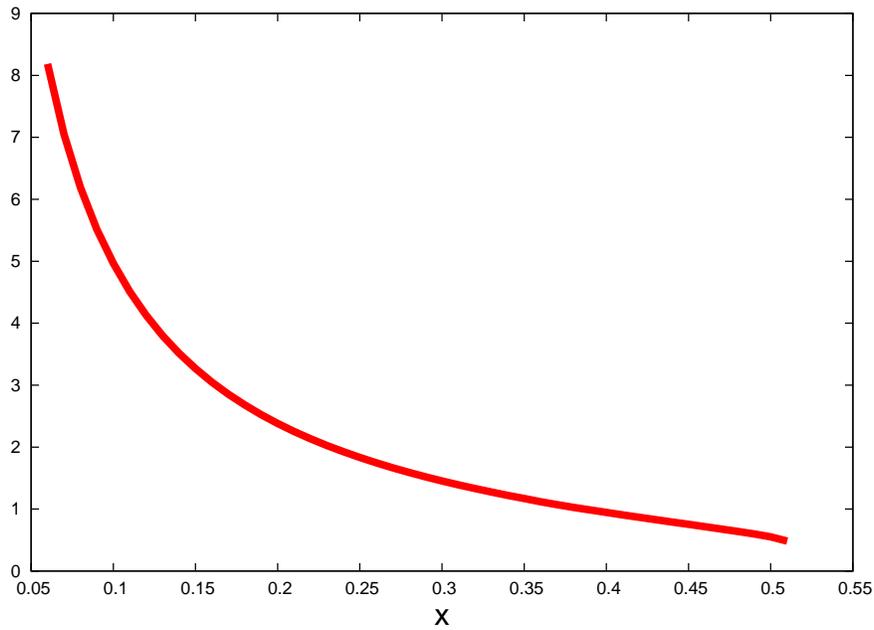, width=120mm}
\caption{\label{2}The ratio $\tilde\sigma/\sigma_0$, Eq.~(\ref{rat}), for the root $A_1(x)$ of the 
equation $G[A]=x$.}
\end{figure}

\begin{figure} 
\epsfig{file=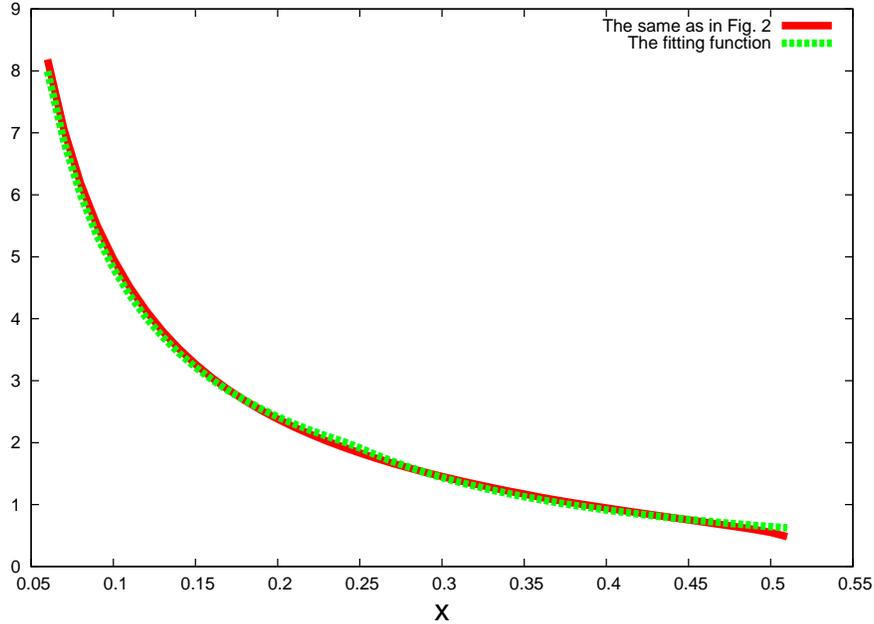, width=120mm}
\caption{\label{4}The ratio $\tilde\sigma/\sigma_0$, Fig.~\ref{2}, and the 
fitting function, Eq.~(\ref{fitF}).}
\end{figure}

Using now the known value of the solid angle in 6 dimensions,
$\Omega_6=\pi^3$, and fixing $N_c=3$, we arrive at a remarkably simple expression:
$$\bigl<\Gamma[A_\mu^a]\bigr>=-\frac{15N_{\rm f}}{(4\pi)^2}\cdot\frac{1}{8}
\int_0^\infty\frac{ds}{s}\cdot\frac{{\rm e}^{-m^2s}}{\tilde\sigma^6}
\int_0^{1/s}
dB\frac{B^7}{\left(1+\frac{B^2}{2\tilde\sigma^2}\right)^{7/2}}.$$
Accordingly, the quark condensate reads
$$\bigl<\bar\psi\psi\bigr>=-\frac{15N_{\rm f}}{64\pi^2}\cdot m
\int_0^\infty ds{\,}\frac{{\rm e}^{-m^2s}}{\tilde\sigma^6}\int_0^{1/s}
dB\frac{B^7}{\left(1+\frac{B^2}{2\tilde\sigma^2}\right)^{7/2}}.$$
Integration over $B$ can be performed analytically, and we obtain
\begin{equation}
\label{c1}
\bigl<\bar\psi\psi\bigr>=-\frac{3N_{\rm f}}{4\pi^2}\cdot m\int_0^\infty ds{\,} 
{\rm e}^{-m^2s}{\,}\tilde\sigma^2\cdot f[A(s)],
\end{equation}
where 
$$A(s)\equiv\frac{1}{2\tilde\sigma^2s^2}~~ {\rm and}~~ 
f[A]=\frac{\bigl(\sqrt{1+A}-1\bigr)^4\cdot\bigl(5A+4\sqrt{1+A}+6\bigr)}{(1+A)^{5/2}}.$$
We arrive now at the crucial observation~\cite{bc} that, in the small-mass limit,
$\bigl<\bar\psi\psi\bigr>$ can only be finite if
\begin{equation}
\label{c2}
\tilde\sigma^2\cdot f[A]\rightarrow\frac{\sigma_0^{3/2}}{\sqrt{s}}~~~~ {\rm at}~~~~ s\to\infty,
\end{equation}
where $\sigma_0$ is some constant parameter to be determined. Because of the relation $\tilde\sigma^2=\frac{1}{2As^2}$,
we conclude that, at large $s$, the following equality should hold:
\begin{equation}
\label{eA}
\frac{f[A]}{A}=x,~~~~ {\rm where}~~~~ x\equiv2(\sigma_0s)^{3/2}.
\end{equation}
Now changing $s\to x$ in $\tilde\sigma^2=\frac{1}{2As^2}$,
we obtain
\begin{equation}
\label{rat}
\frac{\tilde\sigma}{\sigma_0}=\sqrt{\frac{2^{1/3}}{x^{4/3}\cdot A(x)}},
\end{equation}
where $A(x)$ is a root of Eq.~(\ref{eA}). The function $G[A]\equiv\frac{f[A]}{A}$ is plotted in Fig.~\ref{1}.
It vanishes at $A=0$ and $A\to\infty$, and has a maximum at $A\simeq 15.36\equiv\bar A$:
$G[\bar A{\,}]\simeq 0.51\equiv G_{\rm max}$. Therefore, for $x<G_{\rm max}$, equation $G[A]=x$ has two 
roots, $A_1(x)\in(0,\bar A{\,})$ and $A_2(x)\in(\bar A,\infty)$,
while for $x>G_{\rm max}$ it has no roots at all. Thus, we consider only $$x<G_{\rm max},$$
and plot in Fig.~\ref{2} 
the ratio $\tilde\sigma/\sigma_0$, Eq.~(\ref{rat}), as a function of $x$, for the physical root 
$A_1(x)$. It turns
out that the root $A_2(x)$ decreases with $x$ so rapidly that the product 
$x^{4/3}A_2(x)$ decreases as well. As a result, $\tilde\sigma/\sigma_0$ increases with $x$, i.e. with $s$,
which is unacceptable from the physical viewpoint. Instead, the root $A_1(x)$ increases with $x$, and so does 
the product $x^{4/3}A_1(x)$, yielding a physically acceptable decrease of the 
$(\tilde\sigma/\sigma_0)$-ratio with $s$.
This ratio, corresponding to $A_1(x)$, can well be fitted by the following function:
\begin{equation}
\label{fitF}
\left(\frac{\tilde\sigma}{\sigma_0}\right)_{\rm fit}=
\left\{\begin{array}{rcl}\frac{0.49}{x^{0.99}}~~~~~~~~~~~~ {\rm at}~~ x<\frac{G_{\rm max}}{2},\\
\frac{0.22}{x^{1.57}}~~ {\rm at}~~ \frac{G_{\rm max}}{2}<x<G_{\rm max}.
\end{array}\right.
\end{equation}
The result of the fit is plotted in Fig.~\ref{4}. Therefore, at large $s$ of interest, we obtain the 
following scaling behavior:
\begin{equation}
\label{fit1}
\left(\frac{\tilde\sigma}{\sigma_0}\right)_{\rm fit}\propto 
\frac{1}{s^{1.57\cdot\frac32}}\simeq\frac{1}{s^{2.36}}.
\end{equation}
In the next subsection, we will show that, for a Wilson-loop ansatz more realistic than that of  Eq.~(\ref{ans12}),
the fall-off of $\tilde\sigma$ with $s$ becomes linear and fit-independent.

The existence of the maximal value $G_{\rm max}$ of $x$ leads to a lower bound for the value of the 
constituent quark mass $m$. Indeed, from Eqs.~(\ref{c1}) and~(\ref{c2}), we obtain for the 
chiral condensate:
$$\bigl<\bar\psi\psi\bigr>=-\frac{3N_{\rm f}}{4\pi^2}\cdot\sigma_0^{3/2}\int_0^{m^2s_{\rm max}}
\frac{dz}{\sqrt{z}}{\rm e}^{-z}\simeq-\frac{3N_{\rm f}}{4\pi^{3/2}}\cdot\sigma_0^{3/2}~~~ {\rm for}~~~ 
m^2s_{\rm max}=\frac{m^2}{\sigma_0}\left(\frac{G_{\rm max}}{2}\right)^{2/3}\gtrsim 1.$$
Equating further this expression to the commonly accepted  
phenomenological value,
$$\bigl<\bar\psi\psi\bigr>=-{\cal N},~~~~ {\cal N}=(250{\,}{\rm MeV})^3,$$
we get 
\begin{equation}
\label{lb}
\sigma_0=\pi\left(\frac{4{\cal N}}{3N_{\rm f}}\right)^{2/3}~~~ {\rm and}~~~
m\gtrsim2\sqrt{\pi}\left(\frac{\cal N}{3N_{\rm f}G_{\rm max}}\right)^{1/3}.
\end{equation}
In particular, for $N_{\rm f}=2$ light flavors, the lower bound for the constituent quark mass is 
$m\gtrsim 610{\,}{\rm MeV}$. That is too big.
The large value of this bound indicates that the purely exponential fall-off of the Wilson loop 
with the minimal area, assumed in Eq.~(\ref{ans12}), 
is too naive and should be improved. Moreover, the purely exponential fall-off of the Wilson loop 
with the minimal area is incompatible 
with the Gaussian fall-off arising for Wilson loops with sizes smaller than the vacuum correlation 
length~\cite{ds} (cf. Appendix~A). It turns out that these two problems can be simultaneously improved.
That will be done in the next subsection.

\subsection{Diminishing the lower bound for the constituent quark mass}
 
\noindent
A natural way to improve on the purely exponential fall-off of the Wilson loop with the minimal area is 
to consider its additional pre-exponential area-dependence. Such a pre-exponent would not change the 
static quark-antiquark potential. As will be shown in Appendix~A, a suitable parametrization of a Wilson loop,
which, for a constant string tension, accommodates the area dependence for both small and large loops is 
given by the formula
\begin{equation}
\label{mW}
\bigl<W[z_\mu]\bigr>=\frac{N_c}{2^{\alpha-1}\Gamma(\alpha)}\cdot(\tilde\sigma|\Sigma_{\mu\nu}|)^\alpha\cdot
K_\alpha(\tilde\sigma|\Sigma_{\mu\nu}|),
\end{equation}
where $\Gamma(x)$ and $K_\alpha(x)$ stand, respectively, for Gamma- and MacDonald functions, 
and $\alpha>0$ is some parameter. Equation~(\ref{mW}) obeys the normalization condition $\bigl<W[0]\bigr>=N_c$, 
and reproduces the purely exponential fall-off, Eq.~(\ref{ans12}), for $\alpha=\frac12$.
In Appendix~A, we show that, for a constant string tension,  
the best fit of the quadratic area-dependence
is obtained for $\alpha$ in the narrow interval $\alpha\in(1.83,1.9)$. 
In this subsection, by recalculating the chiral condensate
with the $s$-dependent string tension instead of a constant string tension, 
we show that these values of $\alpha$ lead also to the lowering of the constituent quark mass down to $460{\,}{\rm MeV}$.

\begin{figure}
\epsfig{file=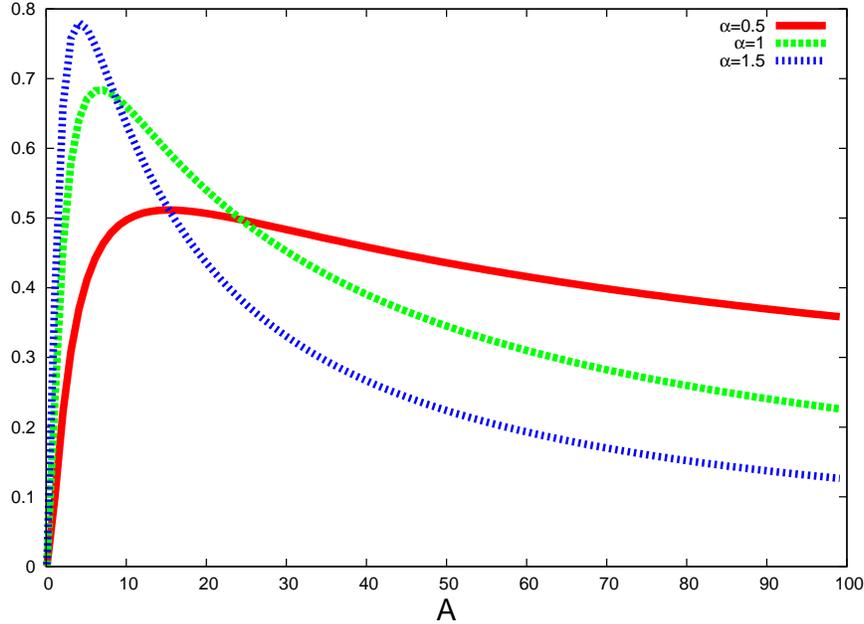, width=120mm}
\caption{\label{5}The function $\tilde G[A,\alpha]=\frac{\tilde f[A,\alpha]}{A}$ for $A\in(0,100)$ and 
$\alpha=0.5$ (cf. Fig.~\ref{1}), $\alpha=1$, and $\alpha=1.5$.}
\end{figure}

The calculation of the condensate can be done by 
using for Eq.~(\ref{mW}) an integral representation
$$\bigl<W[z_\mu]\bigr>=\frac{N_c}{\Gamma(\alpha)}\int_0^\infty d\lambda\lambda^{\alpha-1}
{\rm e}^{-\lambda-
\frac{\tilde\sigma^2}{4\lambda}\Sigma_{\mu\nu}^2},$$
and introducing the field $B_{\mu\nu}$ through the Hubbard--Stratonovich transformation~(\ref{HS18}).
We can then again analytically integrate over $\lambda$, and obtain
[cf. Eq.~(\ref{gg77})]:
\begin{equation}
\label{mW1}
\bigl<W[z_\mu]\bigr>=N_c\cdot\frac{\alpha(\alpha+1)(\alpha+2)}{(2\pi\tilde\sigma^2)^3}\cdot
\left(\prod\limits_{\mu<\nu}^{}\int_{-\infty}^{+\infty}dB_{\mu\nu}\right)
\frac{{\rm e}^{-\frac{i}{2}B_{\mu\nu}\Sigma_{\mu\nu}}}{\left(1+\frac{B_{\mu\nu}^2}{4\tilde\sigma^2}
\right)^{\alpha+3}}.
\end{equation}
We proceed along the lines of the previous subsection to obtain 
the following generalization of Eq.~(\ref{c1}):
$$\bigl<\bar\psi\psi\bigr>=-\frac{3N_{\rm f}}{4\pi^2}\cdot m\int_0^\infty ds{\,} 
{\rm e}^{-m^2s}{\,}\tilde\sigma^2\cdot \tilde f[A(s),\alpha],$$
where 
$$\tilde f[A,\alpha]=\frac{4\{6A(1+A)^\alpha (2+A)+
6[(1+A)^\alpha-1]-(2+\alpha)A[6+(1+\alpha)A(3+\alpha A)]\}}{3(\alpha-1)(1+A)^{\alpha+2}}.$$
In particular, one can check that $\tilde f[A,1/2]=f[A]$, and that, despite the $(\alpha-1)$-term in the 
denominator, $\tilde f$ has a finite limit $\alpha\to 1$:
$$\tilde f[A,\alpha]\to\frac{4[6(1+A)^3\ln(1+A)-A(11A^2+15A+6)]}{3(1+A)^3}~~~ {\rm at}~~~ \alpha\to 1,$$
i.e. $\tilde f$ is a continuous function of $\alpha$. In Fig.~\ref{5}, we plot the function 
$\tilde G[A,\alpha]=\tilde f[A,\alpha]/A$ for three different values of the parameter $\alpha$:
$\alpha=0.5$ (in which case it coincides with the function $G[A]$ from Fig.~\ref{1}), $\alpha=1$, and 
$\alpha=1.5$. We observe a sharpening of the peak with increasing $\alpha$. The  highest peak can be found numerically
to be:
$$\tilde G_{\rm max}[A,\alpha]\to 1.18~~~~ {\rm at}~~~~ \alpha\gg 1.$$ 
The corresponding lower bound for the constituent quark mass, defined by Eq.~(\ref{lb}) with 
$G_{\rm max}$ replaced by $\tilde G_{\rm max}$, is
\begin{equation}
\label{LB}
m_{\rm min}=460{\,}{\rm MeV}.
\end{equation}
Furthermore, the equation $\tilde G[A,\alpha]=x$ again has two roots, one of which, $A_1(x,\alpha)$, 
leads to the decrease of $\tilde\sigma$ with the increase of $x$. 
Because of the sharpening of the peak of $\tilde G[A,\alpha]$ with increasing $\alpha$,
this root behaves as $A_1(x,\alpha)\sim x^{\varepsilon(\alpha)}$, where $\varepsilon(\alpha)\to 0$ at $\alpha\to\infty$. 
The corresponding $x$-dependence of the effective string tension
$\tilde\sigma$ is determined by Eq.~(\ref{rat}) with $A(x)$ replaced by $A_1(x,\alpha\to\infty)$, and reads
$\tilde\sigma\sim x^{-2/3}$. Recalling the definition of $x$ in terms of $s$, Eq.~(\ref{eA}), 
we conclude that  
\begin{equation}
\label{TS}
\tilde\sigma=\frac{c}{s}~~~~ {\rm at}~~~~ \alpha\gg 1,
\end{equation}
where $c$ is constant. 
Thus, the proper-time scaling of $\tilde\sigma$ becomes strictly $1/s$ for $\alpha\gg 1$. 
In practice, as one can see from Fig.~\ref{5}, this scaling holds already for $\alpha\gtrsim 1$, 
in particular for the values $\alpha\in(1.83,1.9)$, which for the constant string tension best fit the quadratic area-dependence of small Wilson loops (cf. Appendix~A).

To find the value of the 
coefficient $c$ in Eq.~(\ref{TS}), we notice that
in the real world the string tension $\sigma_{\rm f}$ (where ``f'' stands for fundamental representation)
is equal to $\sigma_{\rm f}=(440{\,}{\rm MeV})^2$ 
at distances $R<R_{\rm s.b.}$, and vanishes at distances $R>R_{\rm s.b.}$.  Here, 
$R_{\rm s.b.}=\frac{2m}{\sigma_{\rm f}}$ (with ``s.b.'' standing for string-breaking) is the 
distance at which the string joining the quarks breaks by converting its energy to the production of 
another quark-antiquark pair, of mass $2m$. Our model replaces such a step-function behavior of 
the string tension by the function $\tilde\sigma$. Therefore, the normalization coefficient $c$ can be found 
by considering the work necessary to adiabatically 
separate, during the proper time $s$, a quark from an antiquark, from the minimal distance $a$ needed for the string to exist to the maximal distance $R_{\rm s.b.}$ which can be reached without string-breaking.
While in the real world this work is equal to $\sigma_{\rm f}\cdot(R_{\rm s.b.}-a)$, in our 
model it is $\int_a^{R_{\rm s.b.}} dR{\,}\tilde\sigma(R)$, 
and $c$ can be found from the equality of these two expressions.
To calculate the last integral, we notice that, during the proper time $s$, a quark going away from an antiquark separates from it by the distance $R=ms$. Therefore, $\tilde\sigma(R)=c\frac{m}{R}$, and we obtain
$$c=\frac{2m-\sigma_{\rm f}a}{m\cdot\ln\left(\frac{2m}{\sigma_{\rm f}a}\right)}.$$
For the value $a\simeq0.34{\,}{\rm fm}$ in SU(3)-theory with quarks~\cite{wq}, and $m=460{\,}{\rm MeV}$,
Eq.~(\ref{LB}), we get an estimate
$$c\simeq1.26.$$

Finally, we compare with $R_{\rm s.b.}$ the maximal distance $ms_{\rm max}$ achievable by a condensed 
quark in our model. The value of $s_{\rm max}$ can be obtained from the formula $x=2(\sigma_0s)^{3/2}$,
Eq.~(\ref{eA}), where $\sigma_0$ is given by 
Eq.~(\ref{lb}), and the maximal value of the variable $x$ is either $G_{\rm max}$ or $\tilde G_{\rm max}$,
depending on the ansatz for the Wilson loop. For instance, for $x_{\rm max}=G_{\rm max}$, we have
$$s_{\rm max}=\frac{1}{4\pi}\left(\frac{3N_{\rm f}G_{\rm max}}{{\cal N}}\right)^{2/3}.$$
Furthermore, because of the common factor of $m$ in both $R_{\rm s.b.}=2m/\sigma_{\rm f}$ 
and $ms_{\rm max}$, it amounts 
to comparing $2/\sigma_{\rm f}$ and $s_{\rm max}$. We obtain
$$\frac{2/\sigma_{\rm f}}{s_{\rm max}}\simeq
\left\{\begin{array}{rcl}3.85~~ {\rm for}~~ x_{\rm max}=G_{\rm max},\\
2.20~~ {\rm for}~~ x_{\rm max}=\tilde G_{\rm max}.
\end{array}\right.
$$
Thus, at the distances which a quark in our model can achieve, {\it the string does not break}.
We proceed now to a discussion of the physical meaning and applications of the obtained scaling
of the effective string tension, Eq.~(\ref{TS}).

\section{Hausdorff dimension of a light-quark trajectory. A model of the deconfinement phase 
transition in SU(3) Yang--Mills theory.}

\noindent
In this section, we argue that 
the linear fall-off with large proper times $s$ of the effective string tension 
is associated with a highly fractal behavior of quark trajectories at large distances, and make 
quantitative predictions out of this behavior.
In a constant electric field ${\bf E}$, classical Euclidean trajectories of a particle of 
mass $m$ and electric charge $(-1)$ are circles of radius $R_{\rm inst}=m/E$. (The subscript ``inst'' is 
due to the fact that these classical trajectories are commonly referred to as world-line 
instantons~\cite{aam}.) The path integral for the leading fluctuating trajectory about the 
world-line instanton is Gaussian~\cite{aam}. Therefore, the Hausdorff dimension of such a 
trajectory is the same as for a free Brownian random walk, i.e., 2. That is, $L\sim R_{\rm inst}^2$,
where $L$ is the length of a fluctuating trajectory.
In our case, $E\sim\tilde\sigma\sim1/s$,
therefore $R_{\rm inst}\sim ms$, as a direct consequence of the linear fall-off of the effective string 
tension $\tilde\sigma$ with $s$. Thus, we have 
\begin{equation}
\label{L}
L\sim (ms)^2.
\end{equation}

Let us now take a quark trajectory passing through two different points, $x$ and $y$, and 
consider a {\it free} random walk from $x$ to $y$, which takes place along the trajectory 
during the proper time $s$. The length $L_{xy}$ of this random walk is of the order of $L$, while 
the end-to-end distance $R_{xy}\equiv|x-y|$ for a free random walk
scales with the proper time as $R_{xy}\sim s^{1/2}$.
Therefore, we have from Eq.~(\ref{L}):
$$R_{xy}\sim L_{xy}^{1/4},$$
i.e. the Hausdorff dimension of the fluctuating trajectory of a light quark is 4 instead of 2.
We emphasize once again that this result is a consequence of the linear fall-off with 
proper time $s$
of the effective string tension $\tilde\sigma$ of a light quark.

The behavior $L_{xy}\sim R_{xy}^4$
indicates that trajectories of lighter quarks  are much more crumpled than the
conventional Brownian random walks, for which $L_{xy}\sim R_{xy}^2$. Specifically, trajectories 
with Hausdorff dimension 4 are characteristic of the so-called branched polymers~\cite{bp}.
These are just polymers whose monomer branches are Brownian random walks themselves~\cite{poly}.
Let us study the case of pure SU(3) Yang--Mills theory, where the deconfinement phase transition
is weak first-order.
We will show that such a phase transition can be described by a gluon chain 
connecting static quarks~\cite{chain}, when this chain has Hausdorff dimension 4.
In particular, we will reproduce the value of the critical exponent 
$$\nu=1/3,$$ which characterizes a weak first-order phase transition~\cite{fisher}. 

To this end, we follow the approach of Ref.~\cite{adp}.
In that paper, a gluon-chain model of the quark-antiquark string, based on the {\it Brownian} random walk, was 
proven useful for a description of the {\it second-order} deconfinement phase transition. 
Here, we generalize this model for the case of the random walk with Hausdorff dimension 4.
It amounts to noticing that the length of a random walk is 
$L=a\cdot$(number of steps), where in the present case (number of steps)=$\frac{s^2}{a^4}$.
(In these and subsequent formulae, $a$ is the length of one step, i.e. a mean distance between 
the nearest gluons in the chain.) Thus, 
$$L=\frac{s^2}{a^3}.$$ 
Next, since each segment of the chain between the 
nearest gluons may have its own color, the entropy of the whole chain is $S=\ln \bigl(N_c^{L/a}\bigr)$, where 
$L/a$ is the total number of segments.
The free energy of the chain reads $F_{\rm chain}=\sigma_{\rm f} L-TS$. The partition function 
of the random walk has the form
$${\cal Z}_{\rm r.w.}(R,T)=\sum\limits_{n}^{}\int_0^\infty\frac{ds}{(4\pi s)^2}\exp\left[
-\frac{R^2+(\beta n)^2}{4s}-\frac{F_{\rm chain}}{T}\right],$$
where the first term in the exponent corresponds to the kinetic energy of the walker.
Here, $R$ is a 3-dimensional quark-antiquark separation, $\beta\equiv 1/T$, and the sum over 
winding modes $n$ runs from $-\infty$ to $+\infty$.
This partition function can therefore be written as 
$${\cal Z}_{\rm r.w.}(R,T)=\sum\limits_{n}^{}\int_0^\infty\frac{ds}{(4\pi s)^2}\exp\left[
-\frac{R^2+(\beta n)^2}{4s}-\frac{s^2}{a^3}\left(\beta\sigma_{\rm f}-\frac{\ln N_c}{a}\right)\right].$$
Changing the variable $s\to 1/s$ and carrying out the resulting saddle-point integration, we obtain
$${\cal Z}_{\rm r.w.}(R,T)\simeq\frac14\sqrt{\frac{1}{3\pi^3a}\left(\beta\sigma_{\rm f}-
\frac{\ln N_c}{a}\right)^{1/3}}\times$$
$$\times\sum\limits_{n}^{}\frac{1}{[R^2+(\beta n)^2]^{2/3}}\cdot
\exp\left[-\frac{3}{4a}\left(\beta\sigma_{\rm f}-
\frac{\ln N_c}{a}\right)^{1/3}\cdot[R^2+(\beta n)^2]^{2/3}\right].$$
At large quark-antiquark separations $R$ of interest, only the zeroth winding mode survives, so that
\begin{equation}
\label{Zrw}
{\cal Z}_{\rm r.w.}(R,T)\rightarrow\frac14
\sqrt{\frac{1}{3\pi^3a}\left(\beta\sigma_{\rm f}-
\frac{\ln N_c}{a}\right)^{1/3}}\frac{1}{R^{4/3}}\exp\left[-\frac{3}{4a}\left(\beta\sigma_{\rm f}-
\frac{\ln N_c}{a}\right)^{1/3}R^{4/3}\right]
\end{equation}
at $R\to\infty$. As follows from the exponent in this formula, our random-walk model describes 
a static potential, which grows at large distances as $R^{4/3}$~\footnote{Note that, due to the 
one-dimensionality of the random walk, this model cannot describe simultaneously the linear potential and 
a critical exponent $\nu$ different from 1/2. In other words, for the string tension, the model can 
predict only the mean-field critical behavior.}. That is, $V_T(R)=\gamma(T)R^{4/3}$,
where the dimensionality of the coefficient $\gamma(T)$ is (mass)$^{7/3}$. The critical behavior of 
$\gamma(T)$ defines the sought value of $\nu$. Following the approach of Ref.~\cite{adp}, we deduce this
critical behavior from the full free energy of the system. The latter is the sum of the zero-temperature potential, 
\begin{equation}
\label{V0}
V_{T=0}(R)=\gamma R^{4/3},
\end{equation}
and the free energy of the random walk, so that
$$\left.\gamma(T)=\gamma-\frac{T}{R^{4/3}}\ln
\frac{{\cal Z}_{\rm r.w.}(R,T)}{{\cal Z}_{\rm r.w.}(R,T_0)}\right|_{R\to\infty}.$$
The free energy of the random walk is normalized by the condition ${\cal Z}_{\rm r.w.}(R,T_0)=1$,
where $T_0$ is the temperature below which the random walk is suppressed due to the weakness of 
string fluctuations. (In what follows, we will prove that 
$T_0$ is by a significant factor of 4.4 smaller than the deconfinement critical temperature $T_c$.)
From Eq.~(\ref{Zrw}) we have 
\begin{equation}
\label{gT}
\gamma(T)=\gamma+\frac{3T}{4a}\left[\left(\beta\sigma_{\rm f}-
\frac{\ln N_c}{a}\right)^{1/3}-\left(\beta_0\sigma_{\rm f}-
\frac{\ln N_c}{a}\right)^{1/3}\right],
\end{equation}
thus obtaining the deconfinement critical temperature,
which happens to be the same as for the second-order phase transition~\cite{adp}:
\begin{equation}
\label{Tc}
\left.T_c\right|_{N_c>1}=\frac{\sigma_{\rm f} a}{\ln N_c}.
\end{equation}
The temperature $T_0$ is fixed by the condition $\gamma(T_c)=0$, yielding  
\begin{equation}
\label{T0}
T_0=\frac{1}{\ln N_c}\cdot\frac{\sigma_{\rm f}}{\frac1a+\left(\frac43\frac{\gamma}{\sigma_{\rm f}}\right)^3
(\ln N_c)^2}.
\end{equation}
Thus, we find the following critical behavior:
$$\gamma(T)\to\frac34\left(\frac{T_c^2\ln N_c}{a^4}\right)^{1/3}\cdot (T_c-T)^{1/3}~~~~ {\rm at}~~~ 
T\to T_c,$$
i.e. the critical exponent $\nu$ is equal to 1/3, that corresponds to a 
weak first-order phase transition~\cite{fisher}. 

Furthermore, as follows from Eqs.~(\ref{Tc}) and 
(\ref{T0}), the temperature $T_c$ exceeds the temperature $T_0$ by a factor of 
\begin{equation}
\label{100}
\frac{T_c}{T_0}=1+a\cdot \left(\frac43\frac{\gamma}{\sigma_{\rm f}}\right)^3(\ln N_c)^2.
\end{equation}
To get a numerical value of this expression, we evaluate $\gamma$ in terms of $\sigma_{\rm f}$. This 
can be done by approximating the zero-temperature potential~(\ref{V0}) as 
$V_{T=0}(R)\simeq\gamma\left<R^{1/3}\right>\cdot R$, 
and equating it to the standard linear potential $\sigma_{\rm f} R$. 
Here, we have used the obvious fact that $R^{1/3}$ varies much slower than $R$. The mean value 
$\left<R^{1/3}\right>$ corresponding to the potential~(\ref{V0}) reads
$$\left<R^{1/3}\right>=\frac{\int_0^\infty dR{\,} R^{1/3}
\exp\left[-\gamma R^{4/3}\cdot{\cal T}\right]}{\int_0^\infty dR{\,} 
\exp\left[-\gamma R^{4/3}\cdot{\cal T}\right]},$$
where ${\cal T}$ is the Euclidean time needed to communicate a signal between the two static quarks.
For our evaluation purposes, it suffices to 
set this time to its minimal value, that is ${\cal T}=R$. The corresponding mean value of $R^{1/3}$ then reads 
$$\left<R^{1/3}\right>\simeq\frac{3\cdot\Gamma(4/7)}{7\cdot\Gamma(10/7)}
\cdot\gamma^{-1/7},$$
where $\Gamma(x)$ again stands for the Gamma-function. The matching condition, $\gamma\left<R^{1/3}\right>=\sigma_{\rm f}$,
yields the desired expression of $\gamma$ in terms of $\sigma_{\rm f}$:
$$\gamma=\left[\frac{7\cdot\Gamma(10/7)}{3\cdot\Gamma(4/7)}\cdot\sigma_{\rm f}\right]^{7/6}
\simeq1.4\cdot\sigma_{\rm f}^{7/6}.$$
Plugging it into Eq.~(\ref{100}) and setting in that equation $N_c=3$, $a=1{\,}{\rm GeV}^{-1}$~\cite{wq, gna}
in SU(3) Yang--Mills theory, and 
$\sigma_{\rm f}=(440{\,}{\rm MeV})^2$, 
we obtain
\begin{equation}
\label{tr}
\frac{T_c}{T_0}\simeq4.4.
\end{equation}
Thus, the relative range of temperatures, in which our random-walk model of the deconfinement phase 
transition is applicable, is rather broad.

Finally, let us consider the limiting case $N_c=1$, i.e. an Abelian theory
with confinement based on the condensation of magnetic monopoles. In such theories the role of gluons 
which the chain passes through is played by the off-diagonal gauge bosons.
Upon the spontaneous breaking by an adjoint Higgs field of the initial SU(2) symmetry down to (compact) U(1), 
these bosons become massive~\cite{tp}. Since the original non-Abelian symmetry is SU(2), one expects that the 
phase transition is second order. For a gluon-chain model based on the random walk with Hausdorff dimension 2, 
this is anyhow the case~\cite{adp}, since the phase transition in that model is second order for $N_c>1$ as well. 
Instead, for the model developed above, where for $N_c>1$ the phase transition is weak first order,
the reduction to the second-order phase transition for $N_c=1$ is not a~priori obvious. To demonstrate it,   
we write Eq.~(\ref{gT}) for $N_c=1$ as
$$\gamma(x)=\gamma+\frac{3(\sigma_{\rm f} T_0^2)^{1/3}}{4a}\cdot x^2(1-x),~~~~ {\rm where}~~~~ x\equiv(T/T_0)^{1/3}.$$
The critical value $x_c$ is defined by the equation $\gamma(x_c)=0$, which explicitly reads
\begin{equation}
\label{xc}
x_c^2(x_c-1)=\frac{4a\gamma}{3(\sigma_{\rm f} T_0^2)^{1/3}}.
\end{equation}
To derive the critical behavior of $\gamma(x)$, we Taylor expand this function near $x_c$, by substituting 
$x=x_c-\xi$, where $\xi\ll x_c$. To the first order in $\xi$, this expansion yields
$\gamma(x)\simeq\xi\cdot x_c(3x_c-2)$. We note further that $x_c(3x_c-2)=f'(x_c)$, where the function 
$f(x)\equiv x^2(x-1)$ is positive-definite together with its derivative for any $x>1$. Since Eq.~(\ref{xc}), defining $x_c$, can be written as 
$f(x_c)=\frac{4a\gamma}{3(\sigma_{\rm f} T_0^2)^{1/3}}$, which is also a positive-definite quantity, we conclude 
that $x_c>1$, and hence $f'(x_c)>0$ as well. Therefore, $x_c(3x_c-2)>0$, and 
$$\gamma(x)\propto (x_c-x)\propto (T_c^{1/3}-T^{1/3})$$
where the proportionality coefficients are positive-definite. Representing now $T$ in the vicinity 
of $T_c$ as $T\simeq T_c\bigl(1-\frac{\tau}{T_c}\bigr)$, where $\tau\ll T_c$, we get  
$$T_c^{1/3}-T^{1/3}\simeq\frac{\tau}{3T_c^{2/3}}\propto (T_c-T),$$
where the proportionality coefficient in front of $(T_c-T)$ is positive-definite too. Therefore, we conclude that 
$$\gamma(T)\propto (T_c-T),$$
i.e. the phase transition is second order, with the critical exponent $\nu$ equal to 1.
We have thus demonstrated that the weak first-order phase transition, described by our
model at $N_c>1$, becomes for $N_c=1$ second order, with the universality class of 2-dimensional Ising model.
The {\it second}-order phase transition at $N_c=1$ is due to the insufficient entropy 
generated by the random walk in that 
case, {\it even} for a walk with Hausdorff dimension 4. As a result of this insufficiency, the model from the 
thermodynamical viewpoint becomes equivalent to the conventional model of the deconfinement phase 
transition based on the condensation of long {\it closed} strings~\cite{pol1}. The latter also belongs 
to the 2d-Ising universality class (for a discussion of this issue see~\cite{adp}).

\section{Summary}

\noindent
It is intuitively clear that the 
condensation of quarks is associated with the 
zigzag character of their motion, leading to a decrease with the Schwinger proper time
of their effective string tension. It is also known that 
small Wilson loops of heavy quarks respect the ``area-squared'' law, 
which reproduces the heavy-quark condensate. In this paper, 
using an ansatz for the Wilson loop interpolating between the area law at large distances and the 
``area-squared'' law at small distances, we have shown that the decrease of the effective string tension 
with the Schwinger proper time occurs linearly. This leads to the 
lower bound~(\ref{LB}) for the constituent quark mass in our model, whereas 
the naive use at all distances of the area law alone yields an unacceptably high value of this mass.

The linear fall-off of the effective string tension with the proper time is associated 
with highly fractal trajectories of light quarks. Quantitatively, we have found that these trajectories 
have Hausdorff  dimension 4. That is, when such a trajectory connects two different points, the length 
of a random walk performed along this trajectory grows as the fourth power of the end-to-end distance. Trajectories 
of this kind are characteristic for the so-called branched polymers, each segment of whose is 
a Brownian random walk. We have furthermore used these specific trajectories to develop a gluon-chain 
model of the weak first-order deconfinement phase transition characterized by the 
critical exponent $\nu=1/3$. Owing to the $Z_3$-symmetry universality argument, supported by numerous lattice 
simulations, the deconfinement phase transition in SU(3) Yang--Mills theory is commonly accepted 
to be of this kind.
For the realistic values of the string tension $\sigma_{\rm f}$ and the vacuum correlation length $a$, 
we have demonstrated a relative broadness of the temperature range where our model 
is applicable [cf. Eq.~(\ref{tr})]. Finally, in the limiting $(N_c=1)$-case, the deconfinement phase 
transition described by our random-walk model becomes second order, as it should due to the 
center-group universality argument [i.e. because the compact U(1) group emerges by means of a spontaneous 
breaking of an SU(2) symmetry, and this symmetry forces a second-order phase transition]. 

In conclusion, our paper suggests a parametrization of the Wilson loop 
compatible with both the heavy-quark condensate at small distances and the chiral condensate at large 
distances. It provides a lower bound for the constituent quark mass, and 
quantifies fractal properties of the light-quark trajectories. Furthermore, the paper develops a model of the deconfinement phase transition in SU(3) Yang--Mills theory. It would be very interesting to explore how the obtained $1/s$-scaling of the effective string tension changes with temperature, as this change should eventually lead to the chiral-symmetry restoration.

\section*{Acknowledgments}

\noindent
The work of D.A. has been supported by the 
Centre for Physics of Fundamental Interactions (CFIF) at Instituto Superior T\'ecnico (IST), Lisbon.
In addition, he thanks Holger Gies and Christian Schubert for the very useful correspondences.

\section*{Appendix A. Fixing parameter $\alpha$ by the heavy-quark limit.}

\noindent
In this Appendix, we illustrate how to fix the parameter $\alpha$ entering ansatz~(\ref{mW}).
To this end, we notice that this ansatz can fit the ``area-squared'' law, which describes the nonperturbative contribution to small Wilson loops~\cite{ds}. 
Let us first rederive this law and show how it reproduces 
the known expression~\cite{svz} for the heavy-quark condensate. 

By using the non-Abelian Stokes' theorem and the cumulant expansion, one can write the Wilson loop 
in the fundamental representation as~\cite{ds}
$$\bigl<W[z_\mu]\bigr>\simeq {\,}{\rm tr}{\,}\exp\left[-\frac{1}{2!}\frac{g^2}{4}
\int d\sigma_{\mu\nu}(x)\int d\sigma_{\lambda\rho}(x')\bigl<F_{\mu\nu}^a(x)T^a\Phi_{xx'}F_{\lambda\rho}^b(x')
T^b\Phi_{x'x}\bigr>\right],\eqno(A.1)$$
where the integrations go over the minimal surface bounded by the quark trajectory $z_\mu(\tau)$, and 
$\Phi_{xx'}\equiv{\cal P}\exp\bigl[ig\int_{x'}^{x} T^a A_\mu^a(u) du_\mu\bigr]$ is the phase factor along 
the straight line. Furthermore, the factor of $1/2!$ in Eq.~(A.1) is due to the cumulant expansion, while 
the factor of $1/4$ is due to the (non-Abelian) Stokes' theorem. Now, for surfaces  
fitting into the circle of the size of the vacuum correlation length $a$, the field-strength tensor 
$F_{\mu\nu}^a(x)$ can be treated as constant, and one approximates the correlation function in Eq.~(A.1) as
$$\bigl<F_{\mu\nu}^a(x)T^a\Phi_{xx'}F_{\lambda\rho}^b(x')
T^b\Phi_{x'x}\bigr>\simeq\frac{\hat 1_{N_c\times N_c}}{N_c}\cdot\frac{1}{12}\left(
\delta_{\mu\lambda}\delta_{\nu\rho}-\delta_{\mu\rho}\delta_{\nu\lambda}\right)\cdot {\rm tr}
(T^aT^b)\cdot \bigl<F_{\mu\nu}^a(x)F_{\mu\nu}^b(x)\bigr>.$$
Using this parametrization in Eq.~(A.1) and 
noticing that ${\rm tr} (T^aT^b)=\frac12\delta^{ab}$, one has 
$$\bigl<W[z_\mu]\bigr>\simeq N_c\cdot\exp\left[-\frac{\bigl<G^2\bigr>}{96N_c}\int d\sigma_{\mu\nu}(x)
\int d\sigma_{\mu\nu}(x')\right],\eqno(A.2)$$
where $\bigl<G^2\bigr>\equiv\bigl<g^2(F_{\mu\nu}^a)^2\bigr>$ is the gluon condensate.
In particular, for a flat non-selfintersecting trajectory $z_\mu(\tau)$, the double surface integral reads 
$\int d\sigma_{\mu\nu}(x)\int d\sigma_{\mu\nu}(x')=2S^2$, where $S\equiv S[z_\mu]$ is the area of the flat 
surface bounded by $z_\mu(\tau)$. Thus, one gets the ``area-squared'' law~\cite{ds}
$$\bigl<W[z_\mu]\bigr>\Biggr|_{{\rm flat},{\,}{\rm non-selfintersecting}{\,}z_\mu(\tau)}\simeq N_c\cdot
\exp\left[-\frac{\bigl<G^2\bigr>}{48N_c}\cdot S^2\right].\eqno(A.3)$$

We apply now Eq.~(A.2) to the calculation of the heavy-quark condensate. To this end, we rewrite 
the double surface integral as 
$$\int d\sigma_{\mu\nu}(x)\int d\sigma_{\mu\nu}(x')=-\frac12\int d\sigma_{\mu\nu}(x)
\int d\sigma_{\mu\rho}(x')\partial_\nu^x\partial_\rho^{x'}(x-x')^2,$$
and use the (Abelian) Stokes' theorem to obtain
$$\int d\sigma_{\mu\nu}(x)\int d\sigma_{\mu\nu}(x')=-\frac12\oint dz_\mu \oint dz_\mu'(z-z')^2.$$
One can notice that only the $(zz')$-term in $(z-z')^2$ yields a nonvanishing contribution 
to the integral, so that
$$\int d\sigma_{\mu\nu}(x)\int d\sigma_{\mu\nu}(x')=-\frac12\oint dz_\nu \oint dz_\nu'(-2zz')=
\left(\oint dz_\nu z_\mu\right)^2.$$ 
Inserting this expression into Eq.~(A.2), we can apply the Hubbard--Stratonovich trick, 
Eq.~(\ref{HS18}), to further write the Wilson loop as  
$$\bigl<W[z_\mu]\bigr>\simeq\frac{N_c}{(8\pi C)^3}\cdot\left(\prod\limits_{\mu<\nu}^{}\int_{-\infty}^{+\infty}
dB_{\mu\nu}\right){\rm e}^{-\frac{B_{\mu\nu}^2}{16C}-\frac{i}{2}B_{\mu\nu}\oint dz_\nu z_\mu},\eqno(A.4)$$
where we have denoted for brevity $C\equiv\frac{\bigl<G^2\bigr>}{96N_c}$. The Wilson loop in this form can now be used in the one-loop effective action, Eq.~(\ref{effA}), to obtain
$$\bigl<\Gamma[A_\mu^a]\bigr>=-\frac{2N_c N_{\rm f}}{(8\pi C)^3}\int_0^\infty\frac{ds}{s}{\rm e}^{-m^2s} 
\left(\prod\limits_{\mu<\nu}^{}\int_{-\infty}^{+\infty}
dB_{\mu\nu}\right){\rm e}^{-\frac{B_{\mu\nu}^2}{16C}}\times$$
$$\times\left\{
\int_{P}^{}{\cal D}z_\mu \int_{A}^{}{\cal D}\psi_\mu\exp\left[-\int_0^s d\tau\left(
\frac14\dot z_\mu^2+\frac12\psi_\mu\dot\psi_\mu+\frac{i}{2}B_{\mu\nu}z_\mu\dot z_\nu-
iB_{\mu\nu}\psi_\mu\psi_\nu\right)\right]-\frac{1}{(4\pi s)^2}\right\}.\eqno(A.5)$$
The last path integral coincides with that of Eq.~(\ref{eh}), with the substitution ${\cal F}_{\mu\nu}\to
B_{\mu\nu}$. Therefore, it is given by the Euler--Heisenberg--Schwinger Lagrangian, which 
to the order ${\cal O}(s^2B_{\mu\nu}^2)$ yields for the curly bracket in the last equation:
$$\left\{\cdots\right\}\simeq\frac{1}{(4\pi s)^2}\cdot\frac{s^2}{3}\sum\limits_{\mu<\nu}^{}B_{\mu\nu}^2.$$
Applying now definition~(\ref{one}) and recalling that the solid angle in 6 dimensions is $\Omega_6=\pi^3$,
we obtain the heavy-quark condensate in the form
$$\bigl<\bar\psi\psi\bigr>_{\rm heavy}=-\frac{4N_c N_{\rm f}}{(8\pi C)^3}\cdot\frac{1}{(4\pi)^2}
\cdot\frac{\pi^3}{3}\cdot m\int_0^\infty ds{\,} {\rm e}^{-m^2s}\int_{0}^{1/s} dB B^7{\rm e}^{-
\frac{B^2}{8C}}.$$
Because of the factor ${\rm e}^{-m^2s}$, the essential values of the proper time are $s\lesssim\frac{1}{m^2}$,
therefore $\frac1s\gtrsim m^2$. In particular, in the leading large-$m$ approximation, one can approximate
the upper limit of $1/s$ in the last integral by $+\infty$, thus decoupling the $s$- and the $B$-integrals
from each other. The $B$-integral then reads 
$$\int_{0}^{\infty} dB B^7{\rm e}^{-\frac{B^2}{8C}}=12288{\,}C^4,$$
leading to
$$\bigl<\bar\psi\psi\bigr>_{\rm heavy}=-\frac{2N_c N_{\rm f}C}{\pi^2m}.\eqno(A.6)$$
Finally, recalling that $C=\frac{\bigl<G^2\bigr>}{96N_c}$, and writing $\bigl<G^2\bigr>$ as 
$\bigl<G^2\bigr>=4\pi\alpha_s\bigl<(F_{\mu\nu}^a)^2\bigr>$, where $\alpha_s=\frac{g^2}{4\pi}$, we 
recover the known expression for the heavy-quark condensate~\cite{svz}
$$\bigl<\bar\psi\psi\bigr>_{\rm heavy}\Biggr|_{N_c=3}=-N_{\rm f}\cdot\frac{\alpha_s
\bigl<(F_{\mu\nu}^a)^2\bigr>}{12\pi m}.$$
This calculation proves that the ``area-squared'' law, Eq.~(A.3), for the nonperturbative contribution to a small
Wilson loop is consistent with the large-mass limit of the quark condensate.

To evaluate parameter $\alpha$ in Eq.~(\ref{mW}), we replace in that equation $\tilde\sigma|\Sigma_{\mu\nu}|\to
\sigma_{\rm f}S$. Introducing in the ``area-squared'' law, 
Eq.~(A.3), instead of $S$ a new variable $x\equiv\sigma_{\rm f}S$, we now want to find $\alpha$, which 
provides the best approximation in the formula
$$\exp\left(-\frac{\bigl<G^2\bigr>}{48N_c\sigma_{\rm f}^2}x^2\right)\simeq\frac{1}{2^{\alpha-1}\Gamma(\alpha)}
x^\alpha K_\alpha(x)\eqno(A.7)$$
for $x\lesssim 1$. We notice first of all that, up to the positive-definite coefficient $\frac{1}{2^{\alpha-1}\Gamma(\alpha)}$, the second derivative of the approximating function reads
$$\frac{d^2}{dx^2}\left(x^\alpha K_\alpha(x)\right)
=-2^{\alpha-2}\Gamma(\alpha-1)+\alpha(2\alpha-1)\cdot 2^{-\alpha}
\Gamma(-\alpha)x^{2(\alpha-1)}+{\cal O}(x^{2\alpha})+{\cal O}(x),~~ \alpha\ne 1.\eqno(A.8)$$
Therefore, for $\alpha>1$, the second term on the right-hand side of Eq.~(A.8) is subleading compared to the first term,
which is negative-definite. Rather, for $\alpha<1$, the second term on the right-hand side of Eq.~(A.8) becomes 
the dominant one, and one can check that the expression which defines its sign, that is $(2\alpha-1)\Gamma(-\alpha)$, is negative-definite for $\alpha\in\left(\frac12,1\right)$, while becoming positive-definite for $\alpha<\frac12$. In the special case of $\alpha=1$,  
$\frac{d^2}{dx^2}\left(x^\alpha K_\alpha(x)\right)\bigr|_{\alpha=1}=\ln x+{\cal O}(1)$, that is also negative-definite for $x<1$.
Therefore, we conclude that for any $\alpha>\frac12$, as considered 
in subsection II~C, one has 
$\frac{d^2}{dx^2}\left(x^\alpha K_\alpha(x)\right)<0$, in the same way as for the left-hand side of 
Eq.~(A.7). This finding suggests the simplest, analytic, way of seeking $\alpha$ by comparing the 
leading ${\cal O}(x^2)$-terms on the two sides of Eq.~(A.7). Indeed, one can check that the 
${\cal O}(x^2)$-term on the right-hand side of Eq.~(A.7) is the dominant one for $\alpha>1$, in which case 
$$\frac{1}{2^{\alpha-1}\Gamma(\alpha)}
x^\alpha K_\alpha(x)=1+\frac{x^2}{4(1-\alpha)}+{\cal O}(x^{2\alpha})+{\cal O}(x^3).$$
Comparing this expansion with 
$$\exp\left(-\frac{\bigl<G^2\bigr>}{48N_c\sigma_{\rm f}^2}x^2\right)\simeq1-\frac{\bigl<G^2\bigr>}{48
N_c\sigma_{\rm f}^2}x^2,$$
we find 
$$\alpha=1+\frac{12N_c\sigma_{\rm f}^2}{\bigl<G^2\bigr>},$$
that is indeed larger than 1. Substituting into this formula $N_c=3$, $\bigl<G^2\bigr>=\frac{72}{\pi}
\frac{\sigma_{\rm f}}{a^2}$~\cite{ds}, where the value of the vacuum correlation length in QCD with 
dynamical quarks is $a=1.72{\,}{\rm GeV}^{-1}$~\cite{wq}, we obtain 
$$\alpha\simeq1.90.$$
A very close value 
$$\alpha\simeq1.83$$
can be found by numerically minimizing $\chi^2$ for the difference between the two sides of Eq.~(A.7)
in the relevant range $x\in(0,0.45)$. Here, the maximal value $x_{\rm max}\simeq0.45$ corresponds 
to the maximal radius $a/2$ of a circular contour for which the ``area-squared'' law is still applicable.
Indeed, the area encircled by such a contour is $S=\pi(a/2)^2$, that yields $x_{\rm max}=\sigma_{\rm f}S
\simeq0.45$.

\begin{figure}
\epsfig{file=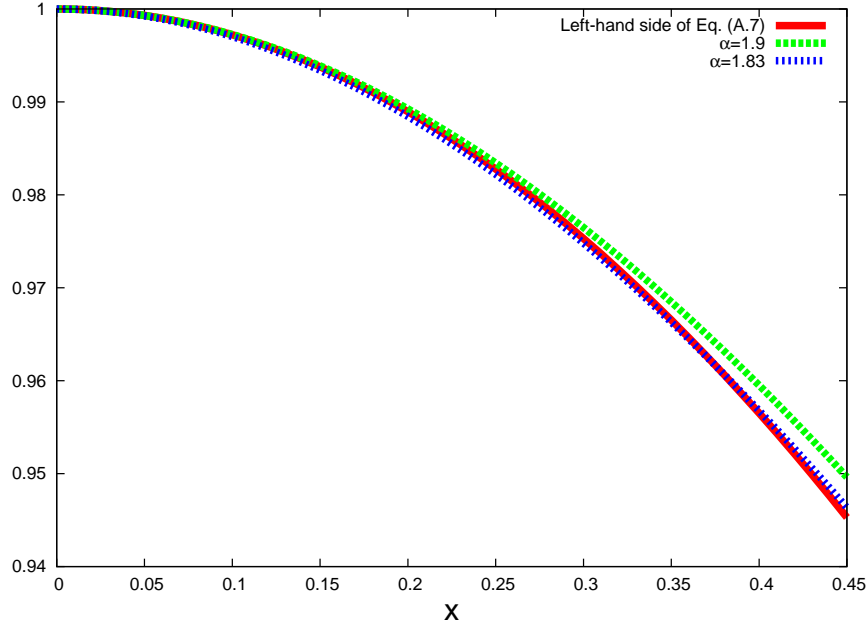, width=120mm}
\caption{\label{59}The left-hand side of Eq.~(A.7) and its right-hand side for $\alpha=1.9$ and $\alpha=1.83$.}
\end{figure}

In Fig.~\ref{59}, we plot the left-hand side of Eq.~(A.7) and its right-hand side for $\alpha=1.9$ and $\alpha=1.83$, and observe a good agreement between these three curves. 
We conclude by noting that, for 
$\alpha\in(1.83,1.9)$, the ansatz 
$$\bigl<W[z_\mu]\bigr>=\frac{N_c}{2^{\alpha-1}\Gamma(\alpha)}\cdot (\sigma_{\rm f}S)^\alpha\cdot
K_\alpha(\sigma_{\rm f}S)$$
approximates the ``area-squared'' law rather well.
At the same time, as we have seen from Fig.~\ref{5}, the linear fall-off of $\tilde\sigma$ with $s$ 
holds with a high accuracy for any $\alpha\gtrsim 1$, in particular for $\alpha$'s from the 
above-mentioned range.

\section*{Appendix B. A relation to the effective quark models.}

\noindent
In this Appendix we derive, as an illustration, the Nambu--Jona-Lasinio action for the simpler case of heavy quarks. To this end, we represent the exponentiated Eq.~(A.5) before the $B_{\mu\nu}$-averaging as 
$$\exp\left\{-2N_cN_{\rm f}\int_0^\infty\frac{ds}{s}{\rm e}^{-m^2s}\times\right.$$
$$\times\left\{
\int_{P}^{}{\cal D}z_\mu \int_{A}^{}{\cal D}\psi_\mu\exp\left[-\int_0^s d\tau\left(
\frac14\dot z_\mu^2+\frac12\psi_\mu\dot\psi_\mu+\frac{i}{2}B_{\mu\nu}z_\mu\dot z_\nu-
iB_{\mu\nu}\psi_\mu\psi_\nu\right)\right]-\right.$$ 
$$\left.\left.-\frac{1}{(4\pi s)^2}\right\}\right\}
=\frac{\det(\gamma_\mu D_\mu+m)}{\det(\gamma_\mu\partial_\mu)}=
\int {\cal D}\psi^a{\cal D}\bar\psi^a\left[{\rm e}^{-\int d^4x\bar\psi^a(\gamma_\mu D_\mu+m)\psi^a}-
{\rm e}^{-\int d^4x\bar\psi^a\gamma_\mu\partial_\mu\psi^a}\right].\eqno(B.1)$$
Here, the Grassmann fields $\psi^a$ and $\bar\psi^a$ represent auxiliary Abelian spin-$1/2$ fermions.
In order to represent the number of quark degrees of freedom, these fields are supplied with additional quantum numbers $a=1,\ldots, N_cN_{\rm f}$. The covariant derivative has the form
$D_\mu=\partial_\mu+iC_\mu$, where $C_\mu(x)=\frac12 x_\nu B_{\nu\mu}$ is the vector-potential 
corresponding to the $x$-independent field-strength tensor $B_{\mu\nu}$. We apply now to both
sides of Eq.~(B.1) the $B_{\mu\nu}$-averaging defined as 
$$\left<\cdots\right>_{B_{\mu\nu}}=\frac{1}{(8\pi C)^3}\left(\prod\limits_{\mu<\nu}^{}\int_{-\infty}^{+\infty}
dB_{\mu\nu}\right){\rm e}^{-\frac{B_{\mu\nu}^2}{16C}}\left(\cdots\right).$$
Since $B_{\mu\nu}^2=2\sum\limits_{\mu<\nu}^{} B_{\mu\nu}^2$, we see that 
this average has the correct normalization, i.e. $\left<1\right>_{B_{\mu\nu}}=1$. Furthermore, owing to Eq.~(A.5), the one-loop approximation adopted throughout the paper yields, for the left-hand side of Eq.~(B.1), the expected expression $\exp\left(\bigl<\Gamma[A_\mu^a]\bigr>\right)$ that is the gluonic-field dependent part of the one-loop QCD partition function. On the right-hand side of Eq.~(B.1), we have
$$\frac{1}{(8\pi C)^3}\left(\prod\limits_{\mu<\nu}^{}\int_{-\infty}^{+\infty}
dB_{\mu\nu}\right){\rm e}^{-\frac{B_{\mu\nu}^2}{16C}}\times$$
$$\times\int {\cal D}\psi^a{\cal D}\bar\psi^a\left[{\rm e}^{-\int d^4x\bar\psi^a\left(\gamma_\mu \partial_\mu
+\frac{i}{2}x_\mu B_{\mu\nu}\gamma_\nu
+m\right)\psi^a}-
{\rm e}^{-\int d^4x\bar\psi^a\gamma_\mu\partial_\mu\psi^a}\right].$$
The second exponent in this expression, which describes the subtracted free part, is of no relevance to 
chiral-symmetry breaking. The $B_{\mu\nu}$-average of the first exponent yields the desired action of a Nambu--Jona-Lasinio--type effective quark model. We note that 
the $B_{\mu\nu}$-integration is the same as in Eq.~(A.4), up to the substitution
$$\oint dz_\nu z_\mu\rightarrow \int d^4x~ x_\mu\bar\psi^a\gamma_\nu\psi^a.\eqno(B.2)$$
For this reason, we can benefit from the already known fact that Eq.~(A.4) is just a representation of Eq.~(A.2). This leads to the following action of a Nambu--Jona-Lasinio--type model:
$$S_{\rm NJL}=\int d^4x~ \bar\psi^a(\gamma_\mu \partial_\mu+m)\psi^a +
\frac{\bigl<G^2\bigr>}{96N_c} \int d^4x\int d^4x'~ x_\mu x_\mu'(\bar\psi_x^a\gamma_\nu\psi_x^a)
(\bar\psi_{x'}^b\gamma_\nu\psi_{x'}^b).$$

The direct product $\gamma_\nu\otimes\gamma_\nu$ can further be disentangled by using the standard Fierz 
transformation. What is important for the present discussion is the {\it quadratic} coordinate dependence of the four-fermion interaction. It has been shown in the first paper of Ref.~\cite{ne} that, in the chiral limit, this type of interaction leads to the chiral condensate 
$\bigl<\bar\psi\psi\bigr>\propto -a\cdot\bigl<g^2{\bf E}^2\bigr>$,
where ${\bf E}$ is the chromo-electric field. The reason why it is the chromo-electric 
rather than the full condensate entering the result is the so-called modified coordinate gauge adopted in~\cite{ne}, 
in which $A_4({\bf 0},x_4)=x_i A_i({\bf x},x_4)=0$. Fixing this gauge, one can eliminate the chromo-magnetic 
part of the full condensate. That part is unlikely to be related to chiral-symmetry breaking, since 
at the deconfinement critical temperature the chiral symmetry is restored, whereas the chromo-magnetic 
condensate not only survives but can even increase further with temperature. At zero temperature, relaxing such a gauge-fixing, one obtains the chiral condensate $\bigl<\bar\psi\psi\bigr>\propto -a\cdot\bigl<G^2\bigr>$.
In this formula, the vacuum correlation length $a$ plays the role of the inverse UV cut-off, which in the heavy-quark limit at issue is replaced by $1/m$. That eventually leads to the QCD sum-rules result for the  heavy-quark condensate, Eq.~(A.6). Remarkably, the $N_{\rm f}$-proportionality of  
$\bigl<\bar\psi\psi\bigr>_{\rm heavy}$ is recovered by the sum over $a$ in the Abelian condensate $\bigl<\bar\psi^a\psi^a\bigr>$.


\end{document}